# Electronic and Magnetic Phase Diagram of Superconductors, SmFeAsO$_{1-x}$F$_x$

Yoichi Kamihara,[1,2] Takatoshi Nomura,[3] Masahiro Hirano,[2,4] Jung Eun Kim,[5] Kenichi Kato,[6] Masaki Takata,[5,6] Yasuhiro Kobayashi,[7,8] Shinji Kitao,[7,8] Satoshi Higashitaniguchi,[7,8] Yoshitaka Yoda,[5,8] Makoto Seto,[7,8,9] and Hideo Hosono[2,3,4]

[1]*JST, TRiP, Materials and Structures Laboratory, Tokyo Institute of Technology Mail Box S2-13, 4259 Nagatsuta, Midori-ku, Yokohama 226-8503, Japan*

[2]*JST, ERATO SORST, Frontier Research Center, Tokyo Institute of Technology, Mail Box S2-13, 4259 Nagatsuta, Midori-ku, Yokohama 226-8503, Japan*

[3]*Materials and Structures Laboratory, Tokyo Institute of Technology, Mail Box R3-1, 4259 Nagatsuta, Midori-ku, Yokohama 226-8503, Japan*

[4]*Frontier Research Center, Tokyo Institute of Technology, Mail Box S2-13, 4259 Nagatsuta, Midori-ku, Yokohama 226-8503, Japan*

[5]*Japan Synchrotron Radiation Research Institute, 1-1-1 Koto, Sayo-cho, Sayo-gun, Hyogo 679-5198, Japan*

[6]*RIKEN SPring-8 Center, 1-1-1, Kouto, Sayo-cho, Sayo-gun, Hyogo 679-5148, Japan*

[7]*Research Reactor Institute, Kyoto University, Kumatori-cho, Sennan-gun, Osaka 590-0494, Japan*

[8]*CREST, JST, Honcho Kawaguchi, Saitama 332-0012,Japan*

[9]*Japan Atomic Energy Agency, 1-1-1 Koto, Sayo-cho, Sayo-gun, Hyogo 679-5148, Japan*

E-mail: kamihara@lucid.msl.titech.ac.jp

**Abstract**

A crystallographic and magnetic phase diagram of SmFeAsO$_{1-x}$F$_x$ is determined as a function of $x$ in terms of temperature based on electrical transport and magnetization, synchrotron powder x-ray diffraction, [57]Fe Mössbauer spectra (MS), and [149]Sm nuclear resonant forward scattering (NRFS) measurements. MS revealed that the magnetic moments of Fe were aligned antiferromagnetically at ~144 K ($T_N$(Fe)). The magnetic moment of Fe (M$_{Fe}$) is estimated to be 0.34 $\mu_B$/Fe at 4.2 K for undoped SmFeAsO; M$_{Fe}$ is quenched in superconducting F-doped SmFeAsO. [149]Sm NRFS spectra revealed that the magnetic moments of Sm start to order antiferromagnetically at 5.6 K (undoped) and 4.4 K ($T_N$(Sm)) ($x = 0.069$). Results clearly indicate that the antiferromagnetic Sm sublattice coexists with the superconducting phase in SmFeAsO$_{1-x}$F$_x$ below $T_N$(Sm), while antiferromagnetic Fe sublattice does not coexist with the superconducting phase.

# 1. Introduction

Recently discovered Fe-based superconductors, $LnFePn(O,F)$ ($Ln$ = La [1]-[4], Ce [4,5], Pr [4], [6]-[8], Nd [4,6,9], Sm [4],[10]-[12], Gd [13,14], Tb, Dy [15,16], Ho [16]; $Pn$ = P, As) [17,18] undergo antiferromagnetic transition at low temperatures when they contain magnetic sublattices of rare earth elements with unfilled 4f shells such as Ce [19], Pr [20,21], Nd [22] and Sm [23]-[25]. Mother compounds of these superconductors crystallize in a tetragonal layered structure, which is composed of an alternate stack of carrier-conducting $FePn$ and carrier-blocking $LnO$ layers (Fig. 1(a)). [26,27] The former layer consists of edge-shared $FePn_4$ tetrahedra with an anti-PbO-type structure, which mainly contributes to the appearance of superconductivity. The Fe elements form another magnetic sublattice in the mother compounds, and are subjected to antiferromagnetic (AF) ordering. [28]

Fe elements having a magnetic moment of 0.35 $\mu_B$/Fe in undoped LaFeAsO, for instance, exhibit AF ordering at temperatures below ~140 K, indicating a Néel temperature of the Fe magnetic sublattice ($T_N$ (Fe)) of ~140 K. [29,30] In addition to the magnetic transition, a crystallographic phase transition from tetragonal to orthorhombic phase takes place at slightly higher temperature than $T_N$(Fe). [31,32] Doping with electrons in the mother compound (undoped LaFeAsO) leads to suppression of both transitions with a reduction of the Fe magnetic moment, and consequently superconductivity appears. [32] This can be performed in SmFeAsO either by doping $F^-$ ions at the O site or formation of $O^{2-}$ vacancies. [4,11] F-doped SmFeAsO exhibits superconductivity with $T_c$ up to ~55 K, which is one of the highest $T_c$ in the Fe-based superconductors so far discovered.

For the present system, temperature dependent heat capacity [24,25] and μ-SR [33] measurements strongly suggest that the Sm magnetic sublattice undergoes AF ordering at low temperatures, although both techniques are not element specific magnetic measurements; Heat capacity and μ-SR measurements are not capable to distinguish main phase's magnetic ordering from impurity's magnetic ordering. In contrast, $^{57}$Fe Mössbauer spectra (MS) and $^{149}$Sm nuclear resonant forward scattering (NRFS) spectroscopy are capable to distinguish using isomer shift values like Knight shift values in nuclear magnetic resonance (NMR). MS and NRFS are also effective in defining element-specific magnetic properties of these compounds making it possible to quantify the magnetic moments of Sm and Fe independently. MS and NRFS spectroscopy provides us information on the magnetic hyperfine field at nuclei position as a function of temperature. [34] Using these techniques, we demonstrated both Néel temperature of the Sm magnetic sublattice ($T_N$ (Sm)) and magnetic moment of the Sm ion for superconducting SmFePO. [10]

For the present system, two different phase diagrams were reported to date. [33,35] Drew *et al* argued that $T_N$(Fe) survives in superconducting SmFeAsO$_{1-x}$F$_x$ with a range $x$ = 0.1 to 0.2. [33] On the other hand, Hess *et al* reported that an apparent coexistence of $T_N$(Fe) and $T_c$ is observed at limited $x$ values within a width of 0.01. [35] The latter report indicates that an apparent coexistence of the superconductivity and AF ordering of Fe (static magnetism) is due mainly to strong inhomogeneous crystallographic phases occurring at the limited $x$ values. Such a discrepancy confuses discussion about the similarity and difference between Fe-based high-$T_c$ superconductors [36] and copper-based high-$T_c$ superconductors. [37] The discrepancy can be dissolved using well characterized samples.

In this study, we prepared polycrystalline SmFeAsO$_{1-x}$F$_x$ ($x$ = 0, 0.005, 0.019, 0.037, 0.040, 0.045, 0.046, 0.060, 0.069, and 0.083). Samples were characterized by x-ray diffraction (XRD), synchrotron x-ray diffraction (SXRD), [38] and resistivity and magnetization measurements, as well as by $^{57}$Fe MS and $^{149}$Sm NRFS spectroscopy, [10,39,40] at various temperatures to define superconducting, magnetic ordering, and crystallographic structure phase transition temperatures. Based on these transition temperatures, we make a phase diagram of SmFeAsO$_{1-x}$F$_x$ in terms of F concentration ($x$) and temperature. The phase diagram we proposed here is closer to that by Hess et al [35]; i.e., long range antiferromagnetic ordering of Fe (a static magnetism) does not persist in the superconducting regime. Such a relation between spin dynamics and superconductivity is common feature among $Ln$FeAsO$_{1-x}$F$_x$ ($Ln$ = La [41], Ce [19], Pr [42], Nd [22] and Sm [35]). Our results indicate that the relation between the static magnetism and $T_c$ of $Ln$FeAsO$_{1-x}$F$_x$ shows similar topology to that of copper-based high-$T_c$ superconductors. [43]

## 2. Experimental

Polycrystalline samples were prepared by a two-step solid-state reaction in a sealed silica tube using dehydrated $Sm_2O_3$ and a mixture of compounds composed of SmAs, $Fe_2As$, and FeAs (SmAs-$Fe_2As$-FeAs powder) as starting materials. The dehydrated $Sm_2O_3$ was prepared by heating commercial $Sm_2O_3$ powder (Rare Metallic Co. Ltd.; 99.99 wt.%) at 1000 °C for 5 h in air. To obtain the SmAs-$Fe_2As$-FeAs powder, Sm (Nilaco; Sm with purity 99.9 wt.%), Fe (Ko-jundo Chemical Laboratory; >99.9 wt.%), and As (Ko-jundo Chemical Laboratory.; 99.9999 wt.%) were mixed in a stoichiometric ratio of 1:3:3 and heated at 850 °C for 10 h in an evacuated silica tube. Then, a 1:1 mixture of the two powders (dehydrated $Sm_2O_3$ and SmAs-$Fe_2As$-FeAs powders) was pressed and heated in a sealed silica tube at 1300 °C for 15 h to obtain a sintered pellet. To prevent the silica tube from collapsing during the reaction, the tube was filled with high-purity Ar gas with a pressure of 0.2 atm at room temperature (RT). All procedures were carried out in an Ar-filled glove box (MIWA Mfg; $O_2$, $H_2O$ < 1 ppm). F doping was performed by replacing part of the $Sm_2O_3$ with a 1:1 mixture of $SmF_3$ (Rare Metallic Co. Ltd.; 99.99 wt.%) and Sm metal in the starting materials.

Crystal structures, including lattice constants of the tetragonal main phase as well as impurity phases of the sintered powders, were examined by powder XRD (Bruker D8 Advance TXS) at RT using CuKα radiation from a rotating anode with the aid of the Rietveld refinement using Code TOPAS3.12. [44] In addition, SXRD measurements were performed at several temperatures ($T$) from 30 to 300 K at the BL02B2 beamline of SPring-8, Japan using a Debye−Scherrer camera with a 286.5 mm camera radius. [38] Two-dimensional Debye−Scherrer images on an imaging plate were obtained by irradiation with monochromatic x-rays with a fixed wavelength of 0.05 nm. For measurements at low temperatures, ground powder samples were put in capillaries and cooled using a dry $N_2$ or He gas-flow cooling device. SXRD patterns ranging from 4° to 73° ($N_2$ gas cooling) or to 53° (He gas cooling) were obtained with a 0.01° step in 2θ, which corresponds to a 0.042 nm or 0.056 nm resolution, respectively. SXRD patterns were then subjected to Rietveld analysis using

Code Rietan2000. [45]

Measurements for dc electrical resistivity ($\rho$) were performed by a four-probe technique using an Au electrode from 1.8 to 300 K. Magnetization measurements were conducted with a vibrating sample magnetometer (Quantum Design; PPMS VSM option) at $T$ = 2.3–300 K. [57]Fe MS were obtained using conventional equipment at $T$ = 4.2–298 K. [149]Sm NRFS spectra were taken at the BL09XU beamline of SPring-8. The MOTIF package was used for NRFS data analysis. [46,47] Details of the MS and NRFS measurements were reported elsewhere. [10,30]

## 3. Results and Discussion

### 3.1 Crystallographic phase characterization

Figure 1(b) shows XRD patterns of SmFeAsO$_{1-x}$F$_x$ [$x$ = 0 (undoped), $x$ = 0.060, and $x$ = 0.069] samples at RT. Almost all diffraction peaks are assigned to those of the tetragonal phase, except for several weak peaks attributable to insulating SmOF, Sm$_2$O$_3$, and normal conducting FeAs [48], indicating that the samples are dominantly composed of the tetragonal phase. Rietveld analysis further reveals that the total amount of impurity phases is less than 7 vol.%.

Figure 2 shows SXRD patterns for the diffraction angle region of $2\theta$ = 27–28°, where (322) reflection of the tetragonal phase was detected. The (322) reflection of the undoped sample starts to split into two at $T$ < 150 K (Fig. 2(a)), whereas that of the F-doped ($x$ = 0.069) sample shows no splitting (Fig. 2(b)). These observations indicate that a symmetry-lowering crystallographic transition takes place in the undoped sample at $T$ < 150 K, and the transition is suppressed in the F-doped sample. The occurrence of crystallographic transition in the undoped sample and suppression in F-doped sample agree with results reported in Ref. [49]. The F-doped sample in the entire temperature region belongs to the tetragonal $P4/nmm$ space group, but those of the undoped sample in the lower temperature region belong to the orthorhombic $Cmma$ space group. [32,49]

### 3.2 Electrical transport properties

Figure 3(a) shows temperature ($T$) dependences of the electrical resistivity ($\rho$) for SmFeAsO$_{1-x}$F$_x$ with various $x$ values, which demonstrate the following; (i) The $\rho-T$ curves show that the $x = 0.037$, 0.040 samples are normal conductors at temperatures down to 1.8 K, whereas the $x = 0.045$, 0.046, 0.060, 0.069, and 0.083 samples are superconductors. For the $x = 0.045$ sample, $\rho$ decreases below our detection limit (Fig. 3(b)). On the other hand, $\rho$ in the $x = 0.037$, 0.040 samples exhibits finite values at 1.8 K above the detection limit, although it decreases rapidly with decreasing temperature. Thus, it is tentatively concluded that the $x = 0.037$, 0.040 samples are normal conductors, and the $x = 0.045$ sample is a superconductor above 1.8 K. (ii) The $\rho-T$ curve of the $x = 0$ sample (undoped) exhibits an anomalous decrease or kink at 143 K ($T_{\text{anom}}$), denoted by an upward arrow which agrees with the onset temperature of the crystallographic transition. (iii) With increasing F content ($x$), $T_{\text{anom}}$ decreases from 143 K for the undoped sample to 91 K for the $x = 0.040$ sample, and eventually the anomaly vanishes for the $x = 0.045$ sample. With a further increase in the $x$ value, it reappears and shifts to a higher temperature, 161 K for $x = 0.046$ and 198 K for $x = 0.083$. The $\delta T_{\text{anom}} / \delta x$ value changes its sign at $x = \sim 0.045$. (iv) The $\rho-T$ curve of the $x = 0.040$ sample shows a clear thermal hysteresis. (v) The $\rho-T$ curves for both the $x = 0.040$ and 0.045 samples show minima ($T_{\text{min}}$), designated by the open triangles, in addition to a kink at $T_{\text{anom}}$.  Residual resistivity ($\rho_{\text{resid}}$) of the samples shows the largest value at $x = 0.040$. Since $x = 0.040$ corresponds to a region where the normal conductor changes to a superconductor in terms of electron doping, the observed singularities imply that crystallographic disorder exists in this concentration region. (See Figure S2 of supplementary data)

*3.3    Macroscopic magnetic properties*

Figure 4 shows magnetic susceptibility ($\chi_{\text{mol}}$) versus $T$ curves for undoped ($x = 0$) and F-doped ($x = 0.069$) samples of SmFeAsO$_{1-x}$F$_x$ as representative examples. The $\chi_{\text{mol}}$ value for the $x = 0.069$ sample starts to decrease at $T \sim 50$ K and shows large negative values with decreasing $T$, reaching $-2.37$ emu at 2.3 K. The superconducting volume fraction estimated from the value is $\sim 75$ vol.% confirming it as a bulk superconductor. Conversely, $\chi_{mol}$ of the undoped sample is positive at RT and increases gradually with decreasing $T$ (Inset (a) of Fig. 4). However, with a further decrease in $T$, it suffers a sharp decrease, exhibiting a maximum at $T = 5.6$ K, and then it increases again very sharply at $T \sim 2.7$ K (Inset (b) of Fig. 4). The positive $\chi_{mol}$ value over the entire temperature range confirms that undoped SmFeAsO is a normal conductor. The maximum at 5.6 K in the $\chi_{mol}-T$ curve is associated with a magnetic transition from paramagnetic to AF phase caused by a Sm magnetic sublattice. [24,25,33] On the other hand, the sharp increase at lower temperature can be explained by two assumptions; One of them is a Curie paramagnetic term due to a magnetic impurity phase and the other one is due to an additional magnetic configuration transition occurring in orthorhombic SmFeAsO phase, known as spin reorientation. [31,32,50] To determine the origin more clearly, further microscopic magnetic studies are essential. Inset (c) of Fig. 4 shows an expanded $\chi_{mol}-T$ curve of $x = 0$ at $T = 130-160$ K, demonstrating a small convex peak at $T = 144$ K, which agrees well with $T_{anom}$. The magnetic transitions associated with the magnetic moment of Fe in $Ln$FeAsO, including LaFeAsO, CeFeAsO, and SmFeAsO, are not clearly detected by macroscopic measurements such as $\chi_{mol}-T$ curves. [5,31]

### 3.4    $^{57}Fe$ Mössbauer spectroscopy

Figure 5 shows $^{57}$Fe Mössbauer spectra (MS) of undoped ($x = 0$) and F-doped ($x = 0.069$) samples in SmFeAsO$_{1-x}$F$_x$ at $T = 298-4.2$ K. Spectra for both samples at RT are composed of a single absorption line indicating the paramagnetic phase of the Fe magnetic sublattice (PM). They were fitted to the singlet pattern. MS of the undoped sample were fitted to the singlet pattern at $T = 150-298$ K. However, with a decrease in $T$, the single absorption line becomes broader and shows a tail of the main absorption line at $T = 150$ K. With further decrease in $T$, they undergo multiple splitting at $T \leq 140$ K, and a sextet split spectrum is observed at 4.2 K, which is a typical behavior of magnetic ordered Fe-compounds. [30,51] If the splitting is caused by an internal magnetic field ($H_{int}$), $H_{int}$ starts to appear at $T \sim 140$ K. (See Figure S3(a) of supplementary data) It increases with decreasing $T$ and reaches 5.16(1) tesla at $T = 4.2$ K, using a conversion factor (CF = 15 tesla/$\mu_B$) [52] corresponding to the magnetic moment of 0.34 $\mu_B$/Fe at 4.2 K. An antiferromagnetic Fe sublattice (AF$_{Fe}$) is clearly observed in undoped SmFeAsO. The second and fifth lines in the sextet split spectrum are enhanced at $T$ = 4.2 K, indicating that magnetic moments of Fe are directed perpendicular to the $c$-axis in the same way as those of LaFeAsO. [30]

*3.5    $^{149}$Sm nuclear resonant forward scattering*

Figure 6 shows time-resolved NRFS spectra for undoped ($x = 0$) and F-doped ($x = 0.069$) samples of SmFeAsO$_{1-x}$F$_x$ at 200, 100, and 4.5 K. The spectra at 200 and 100 K consist of a "dynamical beat" [40], which relates to the effective thickness of the samples, and a quadrupole splitting component. On the other hand, a complex structure is superposed on the dynamical beat in the spectrum at 4.5 K. The structure is attributable to a "quantum beat," which results from hyperfine splitting in the ground and excited states of Sm nuclei. [10,40] The hyperfine splitting is mainly due to the internal magnetic field (H$_{int}$) produced by Sm ions. Thus, the emergence of the quantum beat verifies the AF phase of the Sm magnetic sublattice, and H$_{int}$ of Sm is evaluated to be ~354 tesla at 4.5 K in both the undoped and F-doped samples. Provided that the spin orbit interaction is a dominant but the magnetic exchange and crystal field are negligible small, the magnetic moment is estimated to be ~0.74 $\mu_B$/Sm for a CF = 473 tesla/$\mu_B$ for Sm. [10,34] Figure 7 shows the temperature dependence of NRFS intensity emitted from $^{149}$Sm in the undoped and $x = 0.069$ samples in SmFeAsO$_{1-x}$F$_x$. The NRFS intensities decrease gradually with decreasing temperature and they show sudden increases at 7 K for the undoped sample and at 5 K for the $x = 0.069$ sample (Fig. 7 inset). Further, the intensity for the undoped sample exhibits saturation behavior at $T$ = 5.6 K, denoted by a downward arrow, which agrees with the maximum of χ$_{mol}$ in Fig. 4(b). These behaviors indicate that $T_N$(Sm) of the undoped sample is 5.6 K, and the sudden increase and saturation is most likely associated with antiferromagnetic ordering of Sm ions. Another arrow at $T$ = 4.4 K is the estimated $T_N$(Sm) for $x = 0.069$, assuming the same NRFS intensity at $T_N$(Sm), independent of the F$^-$ concentration. $T_N$(Sm) shifts to lower $T$ in the $x = 0.069$ sample, in which the magnetic moments of Fe spins are quenched. Details of the relation between the NRFS intensity and $T_N(Sm)$ are described in Ref. [10].

*3.6 Phase diagram*

Figure 8 shows a phase diagram for SmFeAsO$_{1-x}$F$_x$ in terms of $x$ and temperature, where $T_{anom}$, $T_c$, $T_{min}$, $T_N(Fe)$, and $T_N(Sm)$ are plotted. Here we assume $T_{anom}$ to be equal to $T_N(Fe)$ for $x \leq 0.040$. SmFeAsO$_{1-x}$F$_x$ are of tetragonal and magnetically disordered phase above the $T_{anom}$-$x$ curve, whereas they are orthorhombic and AF below the $T_{anom}$–$x$ curve for $x \leq 0.040$ and tetragonal and superconducting below the $T_c$–$x$ curve for $x \geq 0.045$. Below $T_N(Sm)$, they are AF independent of the $x$ value, resulting in coexistence of superconductivity and antiferromagnetic Sm sublattice (AF$_{Sm}$) for $x$ larger than 0.045. Although $\rho$ in the $x = 0.037$, 0.040 samples exhibits finite values at 1.8 K, an apparent coexistence of antiferromagnetic Fe sublattice (AF$_{Fe}$) and superconductivity (SC) is observed at limited $x$ values ($x \sim 0.04$) within a width of 0.01. The apparent coexistence of AF$_{Fe}$ and SC indicates crystallographic and/or compositional disorder occurring in the samples. It is indicated that SC does not coexist with AF$_{Fe}$. The F content of $x = 0.045$ may be regarded as the critical F content to induce superconductivity accompanied with suppression of antiferromagnetic ordering of Fe. More homogeneous crystals, which might be obtained using liquid phase reactions, is required to refine the critical F content equal to phase boundary between NC and SC. Our results are similar to that reported in $Ln$FeAsO$_{1-x}$F$_x$ ($Ln =$ La, [41] Ce, [5,19] Pr, [43], Nd, [22] and Sm [35]). Our results reveal that the relation between antiferromagnetic ordering of Fe and $T_c$ of $Ln$FeAsO$_{1-x}$F$_x$ shows similar topology to that of copper-based high-$T_c$ superconductors. [43]

## 4. Conclusions

We prepared polycrystalline SmFeAsO$_{1-x}$F$_x$ ($x = 0$, 0.005, 0.019, 0.037, 0.040, 0.045, 0.046, 0.060, 0.069, and 0.083). Samples were characterized by electrical resistivity, magnetization, and SXRD measurements as well as $^{57}$Fe MS and $^{149}$Sm NRFS spectroscopy at various temperatures to determine transition temperatures for superconductivity, magnetic ordering, and crystallographic structural phases. The results are summarized as follows;

(1) Undoped SmFeAsO undergoes a crystallographic transition from tetragonal to orthorhombic at ~150 K, and an antiferromagnetic transition ($T_N$(Fe) = 144 K) associated with the Fe magnetic sublattice, where the magnetic moment of Fe is estimated to be 0.34 $\mu_B$/Fe at 4.2 K.

(2) Both transition temperatures decrease with an increase in F$^-$ content ($x$) and disappear at $x$ = 0.045. Instead, superconducting transition temperatures appear in $x \geq 0.045$. The transition temperature ($T_c$) and onset temperature ($T_{onset}$) reach 52.5 K and 55.6 K, respectively, at $x$ = 0.083.

(3) An apparent coexistence of $T_N$(Fe) and $T_c$ is observed at rather limited $x$ values ($x \sim 0.04$) within a width of 0.01. We tentatively attribute this apparent coexistence in a very limited $x$-range to crystallographic and/or compositional disorder occurring; i.e., superconductivity does not coexist with magnetic ordered Fe sublattice.

(4) $^{149}$Sm NRFS measurements reveal that Néel temperatures of the Sm magnetic sublattice ($T_N$(Sm)) are located at ~5.6 K for the undoped and at ~4.4 K for a superconducting one ($x$ = 0.069). The magnetic moment of Sm ions is evaluated to be ~0.74 $\mu_B$ at 4.5 K.

Based on these results, a phase diagram for SmFeAsO$_{1-x}$F$_x$ in terms of temperature and $x$ is obtained. The phase diagram provides clear evidence for the coexistence of superconductivity and the antiferromagnetic Sm sublattice in SmFeAsO$_{1-x}$F$_x$ ($x \geq 0.045$) at low temperature, while the antiferromagnetic Fe sublattice does not coexist with superconductivity. $T_N$(Fe) and $T_c$ of $Ln$FeAsO$_{1-x}$F$_x$ show similar topology to that of copper-based high-$T_c$ superconductors.

**Figure captions**

Figure 1. (Color online) (a) Crystal structure of SmFeAsO. The light-grey box represents a unit cell. (b) XRD patterns of SmFeAsO$_{1-x}$F$_x$ [undoped ($x = 0$), $x = 0.060$ and 0.069] at 297 K. The vertical bars at the bottom represent calculated positions of Bragg diffractions of SmFeAsO. The arrows represent Bragg diffractions due to impurity phases (SmOF, Sm$_2$O$_3$, and FeAs).

Figure 2. (Color online) Temperature ($T$) dependence of synchrotron x-ray diffraction (SXRD) patterns of SmFeAsO$_{1-x}$F$_x$ [undoped ($x = 0$) (a), F-doped ($x = 0.069$) (b)]. The vertical bars represent the calculated positions of Bragg diffractions at 30 K. The arrows denote Bragg diffractions of (322) in the tetragonal phases. Temperature dependences of the lattice constants are demonstrated in supplementary data (Fig. S1).

Figure 3. (Color online) Electrical resistivity ($\rho$) of SmFeAsO$_{1-x}$F$_x$ as a function of temperature ($T$). (a) The solid lines (red) and dashed lines (blue) indicate measurements during heating and cooling respectively. Only one sample ($x = 0.040$) shows large thermal hysteresis in the $\rho-T$ curve. The upward arrows (black and blue) indicate temperatures of an anomalous kink in the $\rho-T$ curve ($T_{\text{anom}}$). The downward arrows (red) indicate onset temperatures ($T_{\text{onset}}$) due to the superconducting transition. Both $T_{\text{anom}}$ and $T_{\text{onset}}$ are obtained from the intersection of the tangents to the $\rho-T$ curve in a higher temperature state and halfway through the transition. The downward triangles indicate minima in the $\rho-T$ curve. Broken lines are visual guide indicating $T_{\text{anom}}$ for $x = 0.046, 0.060, 0.069$, and 0.083. The downward triangles indicate minima in the $\rho-T$ curve. (b) $\rho-T$ curves for SmFeAsO$_{1-x}$F$_x$ ($x = 0.037$ and 0.045). The dotted line denotes the detection limit of our measurement.

Figure 4. (Color online) Molar magnetic susceptibility ($\chi_{\text{mol}}$) of SmFeAsO$_{1-x}$F$_x$ [undoped ($x = 0$) and F-doped ($x = 0.069$)] as a function of temperature ($T$). The dashed line near the bottom denotes

perfect diamagnetism ($\chi_{mol} = -3.15$ emu) for SmFeAsO$_{0.931}$F$_{0.069}$. The inset shows expanded $\chi_{mol}-T$ curves in terms of the vertical axis (a), that of the undoped sample from 0 K to 10 K (b), and that from 130 K to 160 K (c). The arrow in (b) denotes a maximum of $\chi_{mol}$ corresponding to the antiferromagnetic transition temperature of the Sm magnetic sublattice ($T_N$(Sm)) (see 3.5). On the other hand, the arrow in (c) indicates an intersection ($T_{kink}$) of the tangent lines (red slashed lines) to the $\chi_{mol}-T$ curves in high and low temperature regions, agreeing with $T_N$(Fe).

Figure 5. (Color online) $^{57}$Fe Mössbauer spectra of SmFeAsO$_{1-x}$F$_x$ [undoped ($x = 0$, left) and F-doped ($x = 0.069$, right)] at several temperatures described in the figure. The solid lines are fitted patterns.

Figure 6. (Color online) Time-resolved nuclear resonant forward scattering (NRFS) spectra (open circles) emitted from $^{149}$Sm in SmFeAsO$_{1-x}$F$_x$ [undoped ($x = 0$) and F-doped ($x = 0.069$)] at three temperatures, as described in the figure. The solid lines are the fitted patterns using parameters listed in supplementary data (Table).

Figure 7. (Color online) Temperature ($T$) dependence of the nuclear resonant forward scattering (NRFS) intensity (I) of $^{149}$Sm in SmFeAsO$_{1-x}$F$_x$ [undoped ($x = 0$) and F-doped ($x = 0.069$)]. The inset shows an expanded graph at $T < 10$ K.

Figure 8. (Color online) Phase diagram of SmFeAsO$_{1-x}$F$_x$ in terms of $x$ and temperature. $T_{onset}$ (open circles), $T_c$ (closed circles), $T_{min}$ (downward triangles), $T_N$ (Sm) (closed squares), and $T_{anom}$ (closed triangles which are almost equal to $T_N$ (Fe) (= $T_{kink}$ (open squares), and open triangles which are not correlated with both of the crystallographic transition temperature and $T_N$(Fe)) are plotted against $x$. The yellow colored area denotes a region in which crystallographic and/or compositional disorder exists.

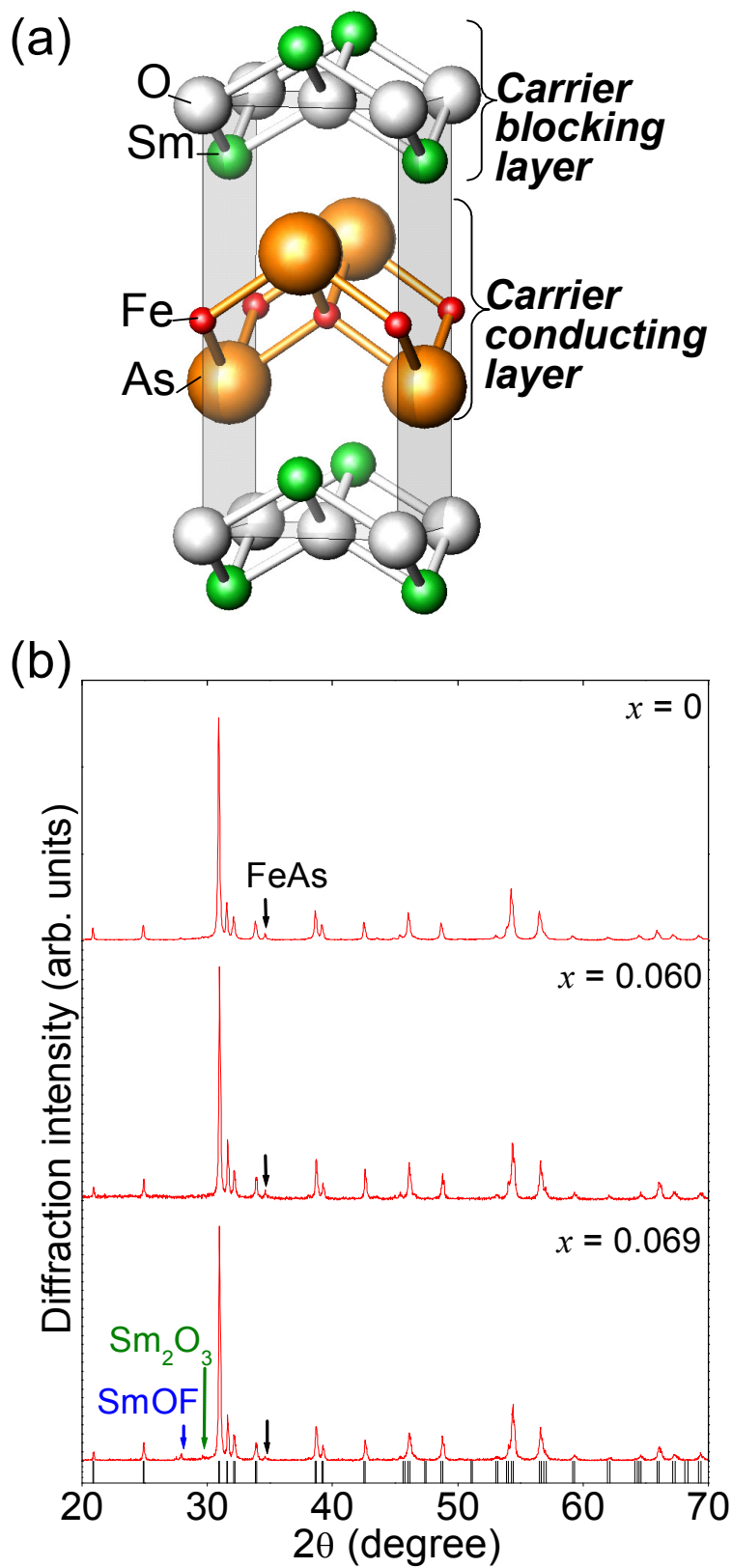

(a)

O
Sm
**Carrier blocking layer**

Fe
As
**Carrier conducting layer**

(b)

$x = 0$

FeAs

$x = 0.060$

$x = 0.069$

Sm$_2$O$_3$

SmOF

Diffraction intensity (arb. units)

$2\theta$ (degree)

Figure 1. Y. Kamihara, *et al*.

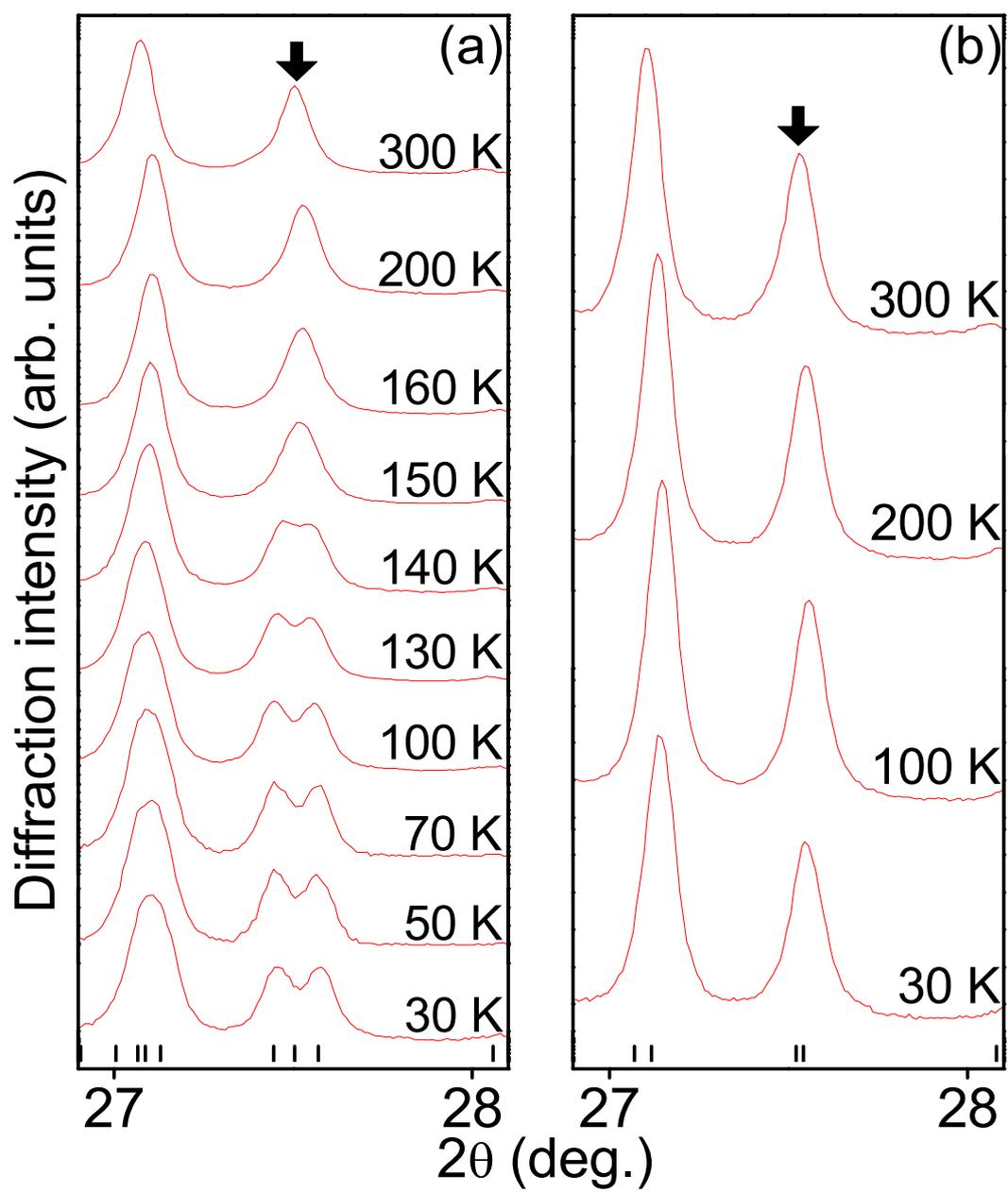

Figure 2. Y. Kamihara, *et al*.

(a)

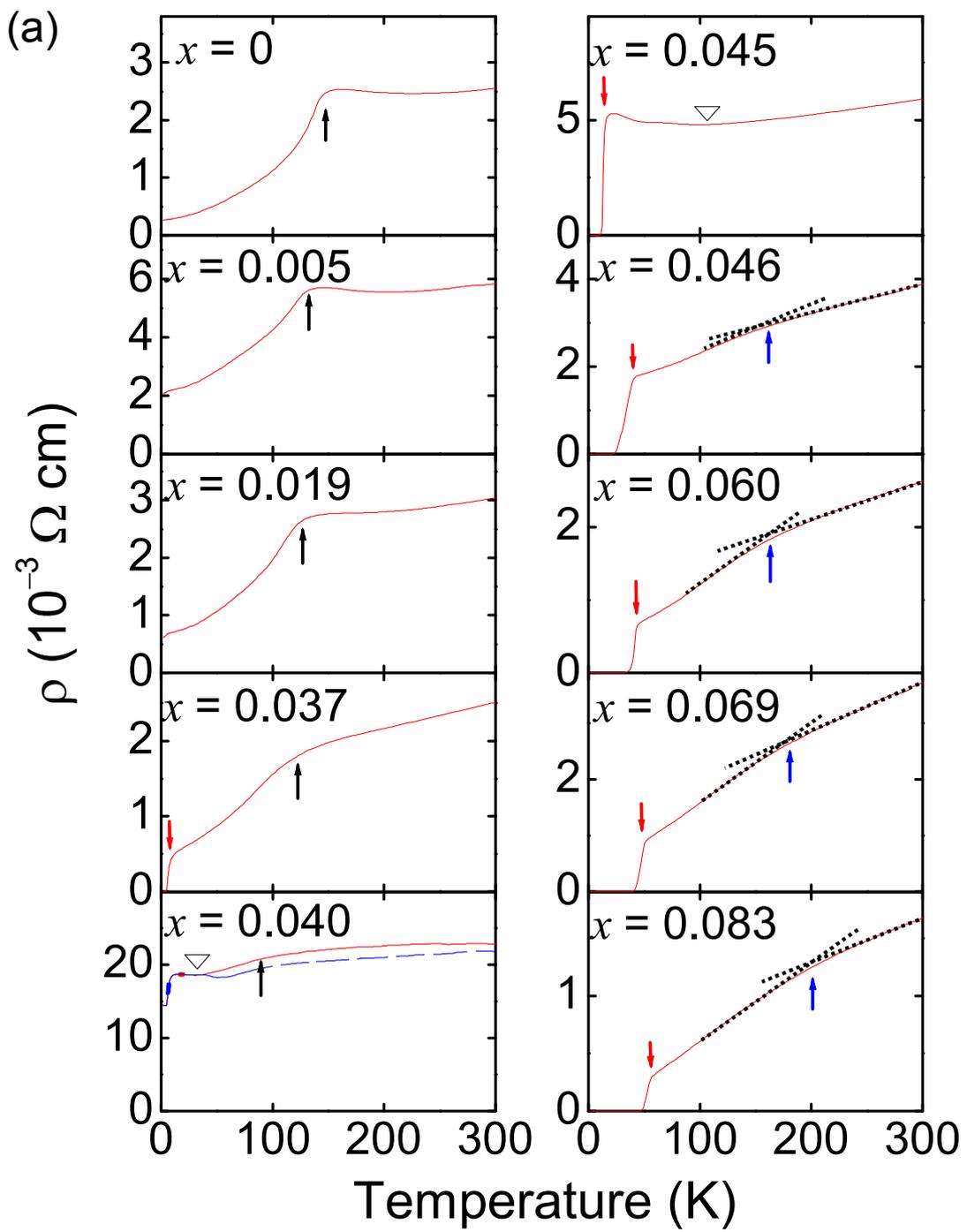

ρ (10⁻³ Ω cm) — $\rho$ ($10^{-3}$ Ω cm)

Temperature (K)

Figure 3(a). Y. Kamihara, *et al*.

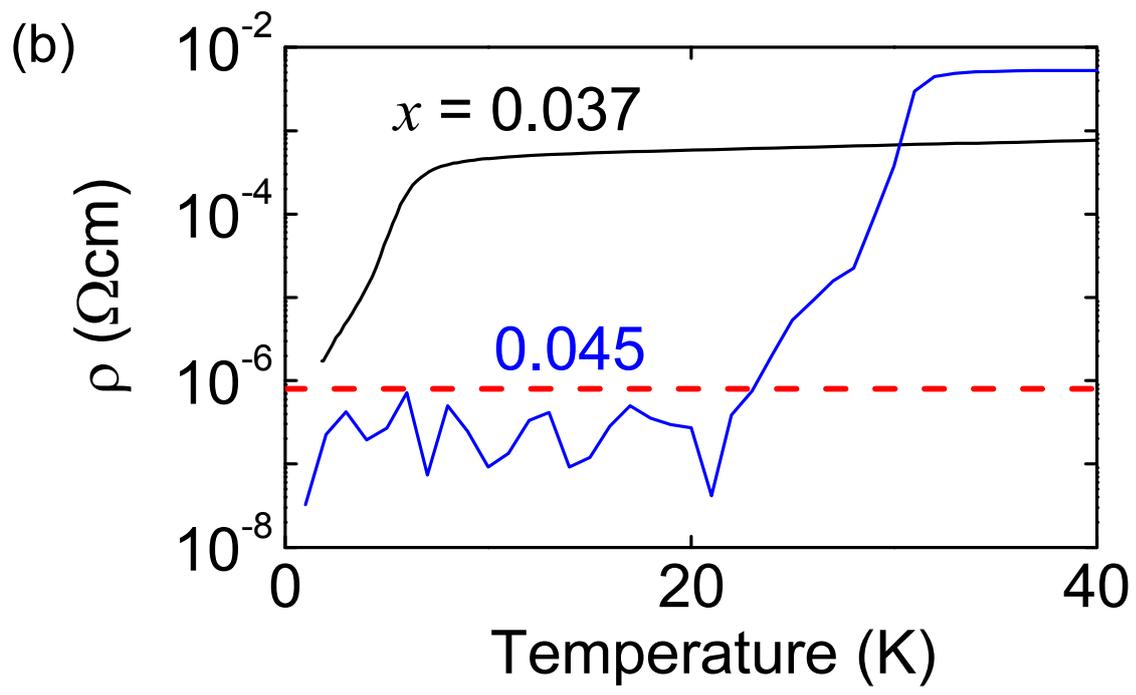

(b)

$x = 0.037$

$0.045$

Figure 3(b). Y. Kamihara, *et al*.

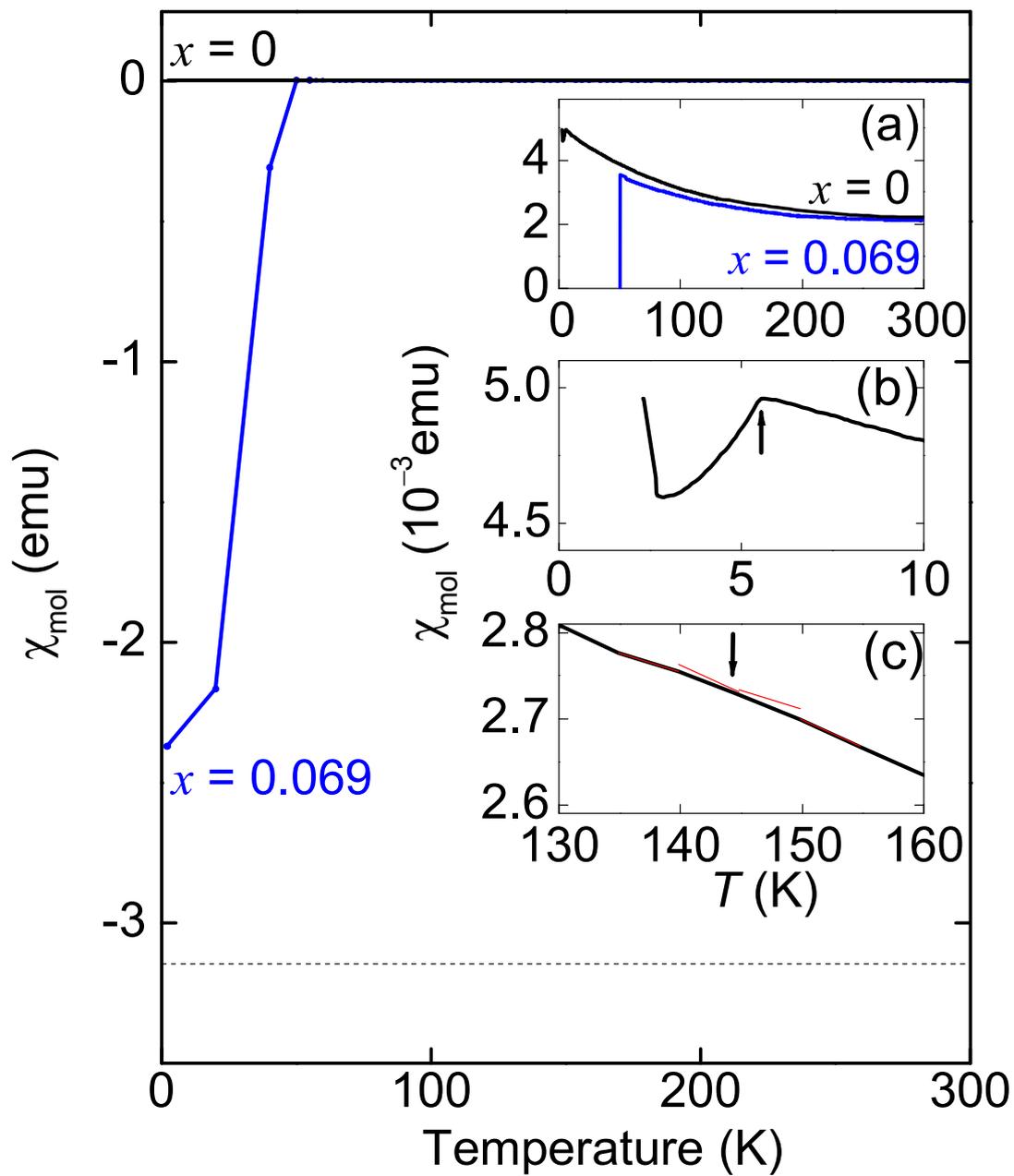

Figure 4. Y. Kamihara, *et al*.

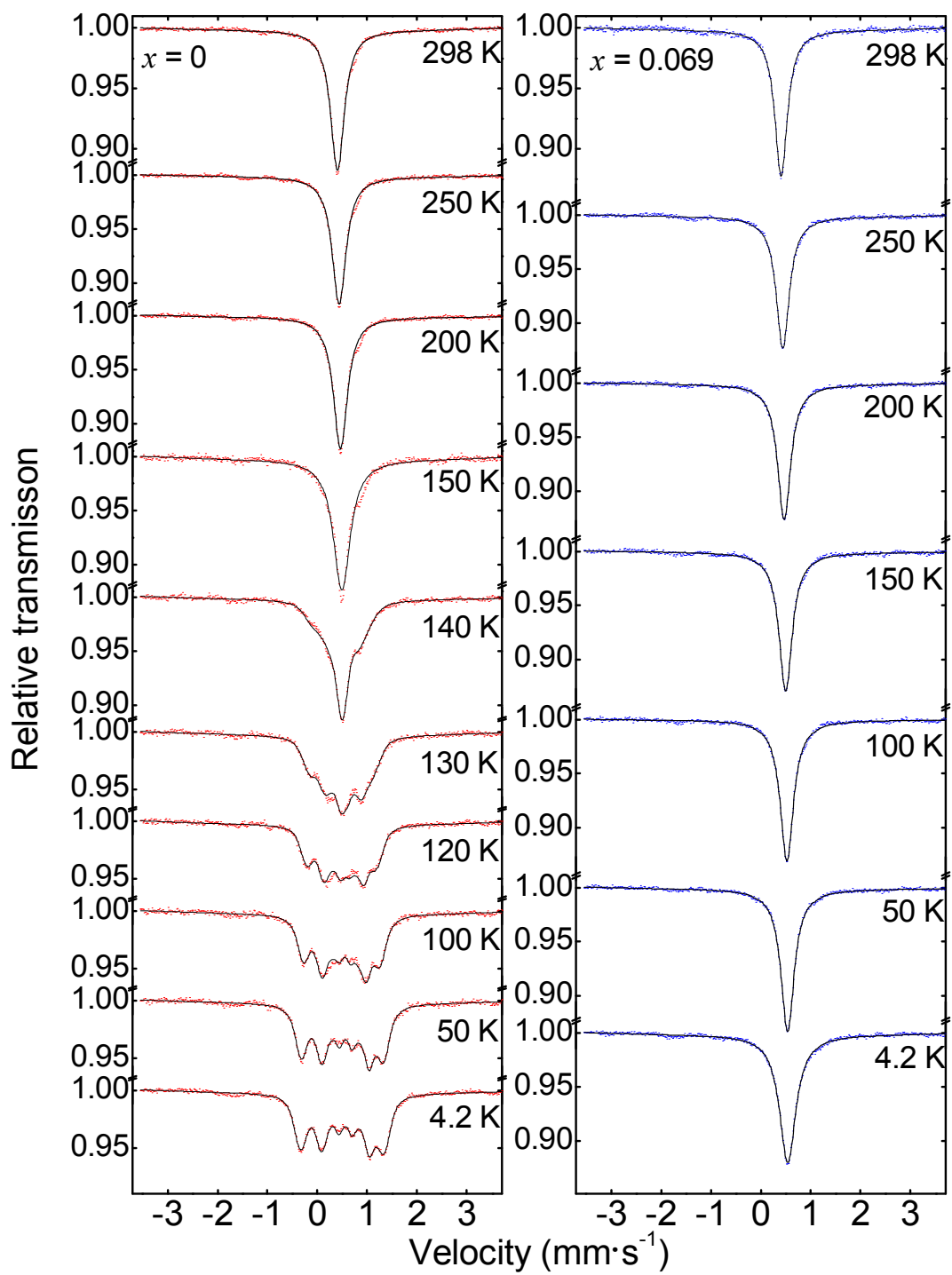

Figure 5. Y. Kamihara, *et al*.

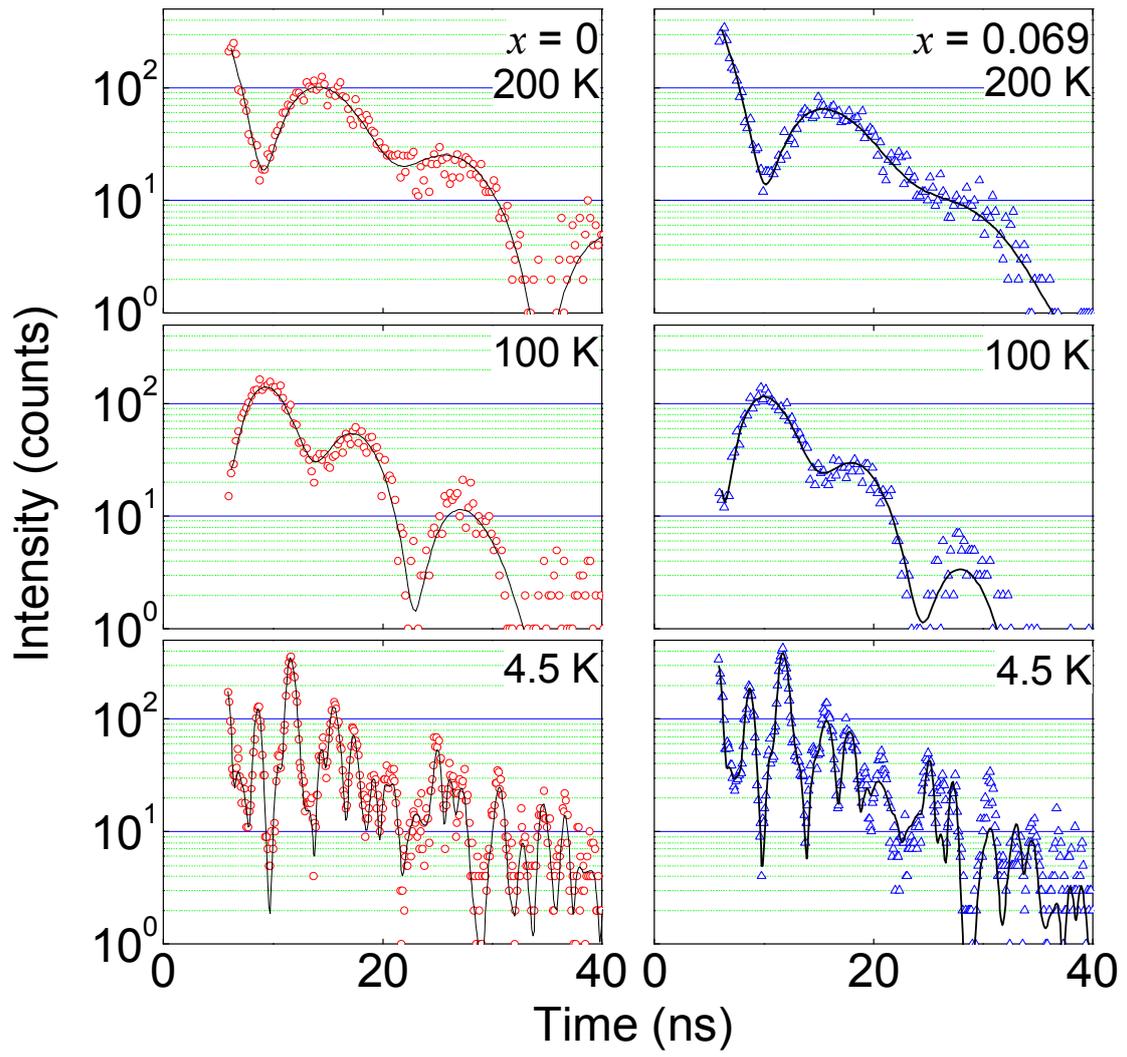

Figure 6. Y. Kamihara, *et al.*

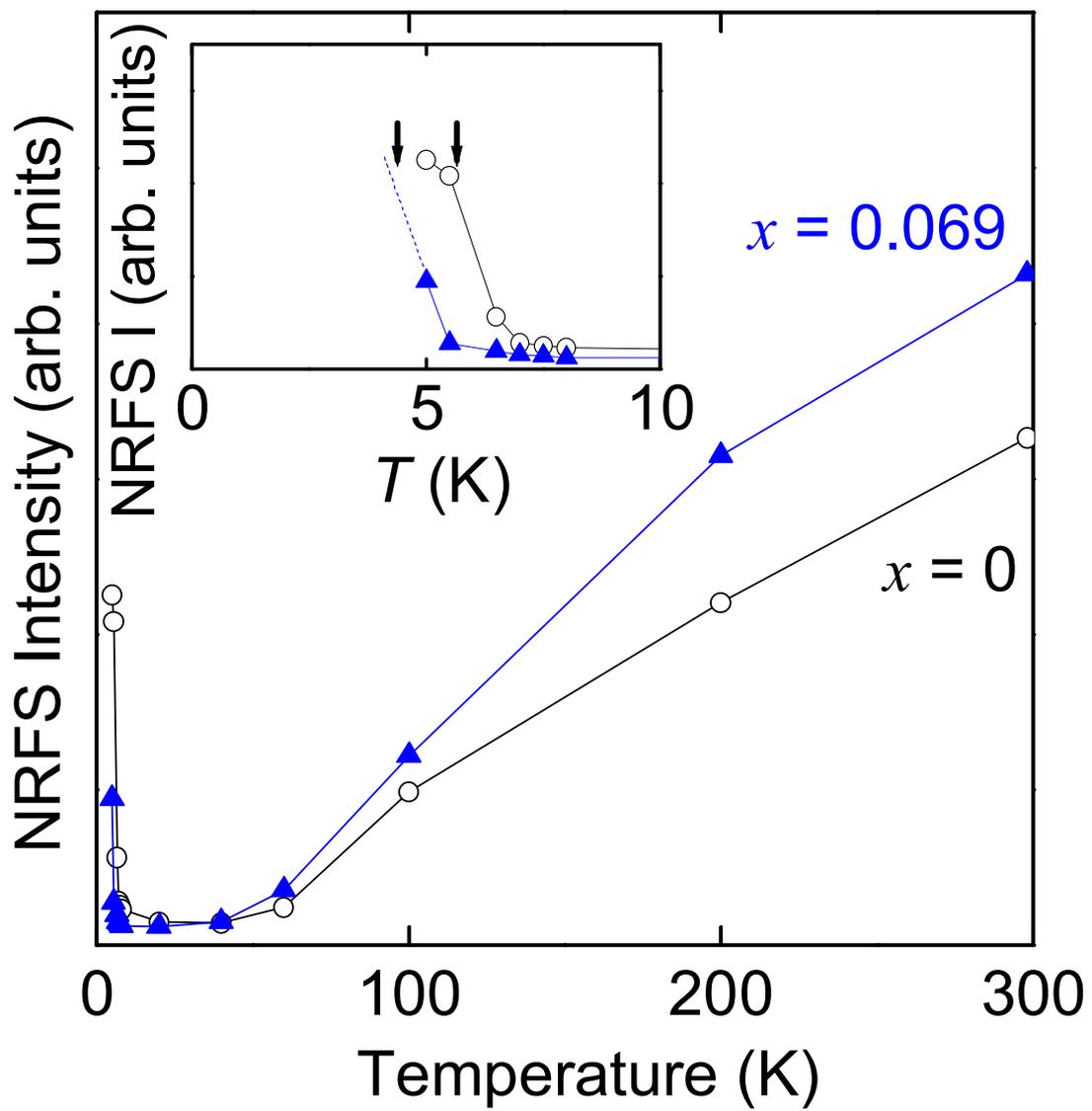

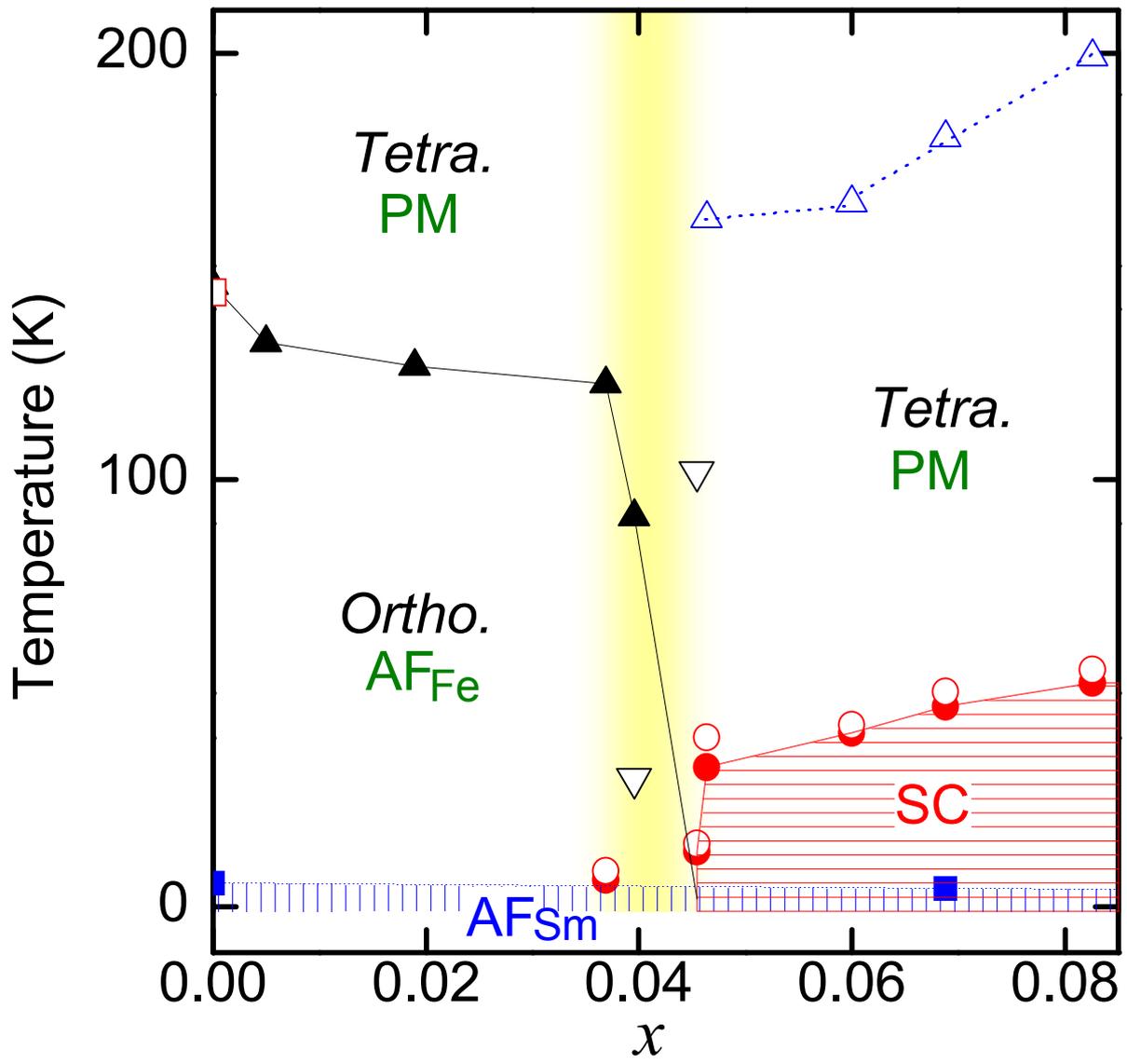

Figure 8. Y. Kamihara, *et al*.